\documentclass[journal,draftcls,onecolumn,12pt,twoside]{IEEEtran}
\ifCLASSINFOpdf
\else
   \usepackage[dvips]{graphicx}
\fi
\usepackage{url}

\usepackage{cite}
\usepackage{amsmath,amssymb,placeins}
\usepackage{algorithmic}
\usepackage{graphicx}
\usepackage{textcomp}
\usepackage[mathscr]{euscript}
\usepackage{lipsum}
\usepackage{cuted}
\usepackage{bm}
\usepackage{lettrine}
\usepackage{multicol}
\usepackage{array}
\usepackage{makecell}
\usepackage{diagbox}
\usepackage[symbol]{footmisc}
\usepackage{tabularx}
\usepackage[usenames]{color}
\usepackage{amsfonts}
\usepackage{latexsym}
\usepackage{subfigure}
\usepackage[format=hang]{subfig}
\usepackage{floatrow}
\usepackage{multirow}
\usepackage{epstopdf}

\newtheorem{theorem}{{{\textit{Theorem}}}}

\newtheorem{lem}{{{\textit{Lemma}}}}

\newtheorem{defn}{{{\textit{Definition}}}}

\newtheorem{remark}{{{\textit{Remark}}}}
\newtheorem{conjecture}{{{\textit{Conjecture}}}}
\newtheorem{example}{{{\textit{Example}}}}

\usepackage{tikz}
\usetikzlibrary{svg.path}

\newcounter{cases}
\newcounter{subcases}[cases]

\usepackage{xcolor}

\begin{document}
 \title{A Construction of Type-II ZCCS for the MC-CDMA System with Low PMEPR}
 \author{Rajen Kumar, Sushant Kumar Jha, Prashant Kumar Srivastava and Sudhan Majhi
 \thanks{Rajen Kumar and  Prashant Kumar Srivastava are with the Department of Mathematics, Indian Institute of Technology Patna, India. e-mail: {\tt rajen\_2021ma04@iitp.ac.in, pksri@iitp.ac.in}. The research presented in this paper was supported in part by the CSIR India, under award letter $09/1023(0034)/2019$-EMR-I.}
 \thanks{Sushant Kumar Jha is with the Department of Electrical Engineering,  Indian Institute of Technology Patna, India. e-mail:{\tt sushant\_2121ee20@iitp.ac.in}}
 \thanks{Sudhan Majhi is with the Department of Electrical Communication Engineering, Indian Institute of Science, Bangalore, India. e-mail: {\tt smajhi@iisc.ac.in}}
 }
 \IEEEpeerreviewmaketitle
 \maketitle
 \begin{abstract}
 In this letter, we propose a novel construction of type-II $Z$-complementary code set (ZCCS) having arbitrary sequence length using the Kronecker product between a complete complementary code (CCC) and mutually orthogonal uni-modular sequences. In this construction, Barker sequences are used to reduce row sequence peak-to-mean envelope power ratio (PMEPR) for some specific lengths sequence and column sequence PMEPR for some specific sizes of codes. The column sequence PMEPR of the proposed type-II ZCCS is upper bounded by a number smaller than $2$. The proposed construction also contributes new lengths of type-II $Z$-complementary pair (ZCP) and type-II $Z$-complementary set (ZCS). Furthermore, the PMEPR of these new type-II ZCPs is also lower than existing type-II ZCPs. % The performance of type-II ZCCS based multi-carrier code-division multiple-access (MC-CDMA) system is provided and compared it with type-I ZCCS. The proposed type-II ZCCS provides a better bit error rate (BER) \textcolor{blue}{when some delay arises} and larger ZCZ width compared to type-I ZCCS. The proposed type-II ZCCS based MC-CDMA also provides more number of codes and better PMEPR compared to CCC and type-I ZCCS.
 \end{abstract}
 \begin{IEEEkeywords}
   Type-II ZCP, Type-II ZCS, CCC, Type-II ZCCS, MC-CDMA, PMEPR.
 \end{IEEEkeywords}
 \section{Introduction}\label{sec:intro}
  \IEEEPARstart{G}{olay}, in his pioneering work in $1961$, proposed a pair of sequences, popularly known as a Golay complementary pair (GCP) \cite{Golay}. The aperiodic auto-correlation function (AACF) value of GCPs is zero except at the zero-shift position. Tseng and Liu \cite{chin} extended the idea of GCPs to a complementary set (CS) that includes two or more constituent sequences with similar AACF value. Paterson \cite{Peter} proposed the construction of CS as a generalization of the construction of GCPs by Davis and Jedwab in \cite{Davis}. Rathinakumar and Chaturvedi \cite{Rat} further extended the idea of Paterson \cite{Peter} to obtain the mutually orthogonal Golay complementary set (MOGCS).
  \par The sequence construction plays a major role in reducing row sequence peak-to-mean envelope power ratio (PMEPR) and column sequence PMEPR. Recently  \cite{Praveen} proposed a novel construction for $2^h$-ary complete complementary code (CCC) of length in the form of $10\cdot 2^m$ and $26 \cdot 2^m$, where $h$ and $m$ are positive integers, where column sequence PMEPR upper bound by $2$ for each code.

In \cite{Fan}, Fan \textit{et~al.} introduced the concept of zero correlation zone (ZCZ) width and proposed Z-complementary pair (ZCP), where AACFs value is zero within this ZCZh. Due to the flexibility in lengths, ZCPs have been widely used over GCPs. ZCPs are known as type-I ZCP, when ZCZ exists around zero shift and type-II ZCP when ZCZ exists towards end shift \cite{LiuO}. Numerous constructions of type-I ZCPs of even length have been provided in \cite{Adhi, Adhi18, LiuE, LiuO, Chen19, Cheng, Chen17, Adhi20}. Similarly, constructions of type-II ZCPs have been provided in \cite{Adhi20, Gu_ZCP_TIT_2021, Zeng_ZCP_WCL_2022, Rajen}. Instead of pair, if a set of sequences follow the property of zero AACF value within the ZCZ, it is known as $Z$-complementary set (ZCS).

A collection of ZCS is called $Z$-complementary code set (ZCCS), if the aperiodic cross-correlation function (ACCF) value of two different ZCS is zero in the ZCZ region, including the zero time shift. The ZCCS is advantageous in the sense that it has a larger set size compared to the CCC with the same number of constituent sequences \cite{Fan}. Type-I ZCCS refers to a set of codes where the ZCZ exists near to zero time shift. Increasing users by $n$ times for a type-I ZCCS based MC-CDMA system over a Quasi synchronous (QS) environment leads to limiting the ZCZ width by $1/n$ times. There is no interference for delays in ZCZ, therefore, a set with larger codes and wider ZCZ is needed to improve the MC-CDMA system. In that direction, how type-I ZCCS is extended from type-1 ZCP, similar way type-II ZCCS could be extended from type-II ZCP.  Moreover, several efforts have been made to tighten the PMEPR for the column sequence of ZCCS or CCC, but to date, the minimum upper bound is tightened by $p$, when elements of code are $p$th root of unity. To the best of the authors' knowledge, there are no constructions that provide larger users, larger delays and better PMEPR of MC-CDMA system then type-I ZCCS.

%whereas for type-II ZCCS, the ZCZ width exists near the end time shift. When the delay spread in wide-band communication exceeds the symbol time, the reflected components of a signal arrive with a temporal delay to the receiver. \textcolor{blue}{In such scenarios, type-II ZCP is more effective than type-I ZCP \cite{W}. A CCC is an extension of the GCP for supporting more users. Likewise, Type-I ZCCS is an extension of type-I ZCP. Similarly, we extended type-II ZCP to type-II ZCCS.} \textcolor{blue}{However, the set size of CCC is limited, which limits the number of users in the MC-CDMA system. So type-I ZCCS is often used to increase the number \textcolor{blue}{of users} for the MC-CDMA system by sacrificing the ZCZ width. So we need a set whose number of codes is \textcolor{blue}{large}, can support more number of user and provide better PMEPR with larger ZCZ width.} To the best of the authors' knowledge, there are no constructions which provide larger users, larger delays  and better PMEPR of MC-CDMA system then type-I ZCCS. on type-II ZCS and \textcolor{blue}{type-II} ZCCS have been reported in the literature.

In this paper, for the first time, we propose type-II ZCCS to increase the number of users and to reduce PMPER for the MC-CDMA system over a large QS system. The construction uses the Kronecker product between a $(K,K,N)$-CCC and $r$ mutually orthogonal uni-modular sequences of length $P$ to produce type-II ZCCS. For a given $(K,K,N)$-CCC, the proposed construction provides type-II $(rK,K,NP-P+1,NP)$-ZCCS. Furthermore, for the fixed value of the constituent sequence, the proposed type-II ZCCS provides a larger set size compared to that of CCC and type-I ZCCS. We introduce an idea of the Barker sequences to reduce the PMEPR of a code. The column sequence PMEPR of type-II ZCCS is upper bounded by a number smaller than $2$, which is lower than the bounds reported in existing literature for a code. The proposed construction also provide type-II ZCPs with new lengths compared to \cite{Zeng_ZCP_WCL_2022} and \cite{Rajen} and provides lower PMEPR compared to \cite{Rajen}. It also contributes type-II ZCS. The performance of type-II ZCCS based QS MC-CDMA system is compared with type-I ZCCS. The proposed type-II ZCCS provides a better bit error rate (BER) and larger ZCZ width compared to existing type-I ZCCS.

The rest of the paper is stated as follows. Section \ref{sec:pre} contains the introduction and notations. Section \ref{sec:prop} describes the proposed construction. In Section \ref{sec:system}, we provided a system model for type-II ZCCS based MC-CDMA. In Section \ref{sec:comp}, the proposed construction has been compared with some of the existing work. Finally, Section \ref{sec:conc} includes the final observations. 
\section{Preliminaries}\label{sec:pre}
This section introduces the fundamental notations, definitions, and lemmas utilized in this letter.
%\vspace{-0.6cm}
\subsection{Basic Notations}
A sequence/code is called $p$-ary if each element satisfies $x^p-1=0$ (for minimum value of $p$), where $p\ge 2$. Further, if $p=2$, it is called binary sequence/code.
\begin{defn}[\cite{Sarkar_Psu}]
	Let $\mathbf{a}=(a_0,a_1,\hdots,a_{N-1})$ and $\mathbf{a'}=(a'_0,a'_1,$$\hdots,$$a'_{N-1})$ be two complex-valued sequences of length $N$. The AACF value between $\mathbf{a}$ and $\mathbf{a'}$ is defined as
	\begin{equation}
	    \rho(\mathbf{a}, \mathbf{a'})(\tau)=\left\{\begin{array}{ll}
		\sum_{i=0}^{N-1-\tau} a_{i+\tau} a_{i}^{'*}, & 0 \leq \tau<N, \\
		\sum_{i=0}^{N+\tau-1} a_{i} a_{i-\tau}^{'*}, & -N<\tau<0, \\
		0, & \text { otherwise, }
	\end{array}\right.
	\end{equation}
	where $a_i^{'*}$ is complex conjugate of $a'_i$. When $\mathbf{a}=\mathbf{a'}$, $\rho(\mathbf{a},\mathbf{a'})(\tau)$ is called AACF of $\mathbf{a}$ and is denoted as $\rho(\mathbf{a})(\tau)$.
\end{defn}
\begin{lem}\label{pac}
	Let $\mathbf{a}$ and $\mathbf{a'}$ be two complex-valued sequences of identical length $N$, where $\mathbf{a'}=e^{\sqrt{-1}\theta}\mathbf{a}$ and $\theta\in [0,2\pi)$. Then
		$\rho(\mathbf{a'})(\tau)=\rho(\mathbf{a})(\tau).$
\end{lem}
\begin{lem}\label{mac}
	Let $\mathbf{a}$ and $\mathbf{a'}$ be two $p$-ary complex-valued sequences of identical length $N$, where $a_i=\zeta^ia'_i$, where $\zeta=e^{\sqrt{-1}\frac{2 \pi}{p}}$. Then
		$\rho(\mathbf{a'})(\tau)=\rho(\mathbf{a})(\tau).$
\end{lem}
\begin{defn}
	Let $\mathbf{C}=\left\{{C}_{1},{C}_{2}, \ldots, {C}_{K}\right\}$ be a set of $K$ matrices (codes), each of having order $M \times N$. Here $C_{k}$ is defined as
 %\vspace{-0.35cm}
	$$
	C_{k}=\left[\begin{array}{c}
		\mathbf{c}_{1}^{k}\\ \mathbf{c}_{2}^{k}\\ \vdots\\\mathbf{c}_{M}^{k}
	\end{array}\right]_{M \times N},
	$$
	where $\mathbf{c}_{\nu}^{k}(1\leq \nu \leq M ,1 \leq k \leq K)$ is the $\nu$-th row sequence of $C_{k}$.\\
	Let $C_{k_{1}}$ and $C_{k_{2}} $ be any two matrices in $\mathbf{C}$, then ACCF of $C_{k_{1}}$ and $C_{k_{2}}$ is defined by
	\begin{equation}
		\rho\left({C}_{k_{1}},{C}_{k_{2}}\right)(\tau)=\sum_{\nu=0}^{M-1} \rho\left(\mathbf{c}_{\nu}^{k_{1}}, \mathbf{c}_{\nu}^{k_{2}}\right)(\tau).
	\end{equation}
When ${C}_{k_1}={C}_{k_2}={C}_{k}$, $\rho(C_{k_1},C_{k_2})(\tau)$ is called AACF of ${C}_{k}$ and is denoted as $\rho(C_{k})(\tau)$.
\end{defn}
\begin{defn}\label{def:ZCCS2}
Let $\mathbf{Z}=\left\{{Z}_{1},{Z}_{2}, \ldots, {Z}_{K}\right\}$ be a set of $K$ matrices (codes), each of having order $M \times N$. $\mathbf{Z}$ is called a type-II $(K,M,Z,N)$-ZCCS with ZCZ width $Z$, if it satisfies the following properties
 %\vspace{-0.35cm}
	\begin{equation}\label{eq:def_ZCCS}
		\rho\left(Z_{k_{1}}, Z_{k_{2}}\right)(\tau)=\left\{\begin{array}{ll}
			NM, & \tau=0, k_{1}=k_{2}, \\
			0, &  \tau=0, k_{1} \neq k_{2},\\
			0, & N-Z<|\tau|<N.
		\end{array}\right.
	\end{equation}
\end{defn}
We are proposing a conjecture on the number of codes based on the maximum number of codes with respect to the number of the constituent sequences, length and ZCZ width.
\begin{conjecture}[Rajen bound]\label{Conj:code}
For type-II $(K,M,Z,N)$-ZCCS, the number of codes $K$ is bounded by
\begin{equation}
    K\le M(N-Z+1).
\end{equation}
\end{conjecture}
Each code $Z_k\in \mathbf{Z}$ is called a type-II ZCS. It is known as type-II ZCP, when the number of rows in a type-II ZCS is $2$. When $Z=N$, a type-II ZCCS is called MOGCS and when $K=M$, a MOGCS is called $(K,K,N)$-CCC. If ACCF values of two codes are zero for all time shifts, these are called uncorrelated codes.
	\begin{lem}[\cite{Jin_CCC_kron_DFT_2008}]\label{clg}
	Let $\mathbf{a}=(a_0,a_1,\ldots,a_{N-1})$ and $\mathbf{a^\prime}=(a^\prime_0,a^\prime_1,\ldots,a^\prime_{N-1})$ of length $N$ and $\mathbf{b}=(b_0,b_1,\ldots, b_{P-1})$ and $\mathbf{b'}=(b'_0,b'_1,\ldots, b'_{P-1})$ be uni-modular sequences (each element of the sequence is of modulus $1$) of length $P$. Let $\otimes$ be the Kronecker product. Then $\mathbf{c}=\mathbf{a}\otimes \mathbf{b}$ and $\mathbf{c^\prime}=\mathbf{a^\prime}\otimes \mathbf{b'}$ are sequences of length $NP$ and ACCF of $\mathbf{c}$ and $\mathbf{c}^\prime$ depends on ACCF of $(\mathbf{a},\mathbf{a^\prime})$ and  $(\mathbf{b},\mathbf{b'})$, as
	\begin{equation}
	\begin{aligned}
	      \rho(\mathbf{c,c^\prime})(Pj+k)=&\rho(\mathbf{a,a^\prime})(j)\rho(\mathbf{b},\mathbf{b'})(k)\\&+\rho(\mathbf{a,a^\prime})(j+1)\rho(\mathbf{b},\mathbf{b'})(-P+k),
	\end{aligned}
	\end{equation}
 where $0\le k < P$ and $-N<  j< N$.
\end{lem}
\begin{IEEEproof}
In \textit{Lemma} \ref{clg}, if we consider $\mathbf{a}=\mathbf{a^\prime}$ then $\mathbf{c}=\mathbf{c^\prime}$ and ACCF of $\mathbf{a},\mathbf{a^\prime}$ and $\mathbf{c},\mathbf{c^\prime}$ become AACF of $\mathbf{a}$ and $\mathbf{c}$. From \textit{Lemma} \ref{clg}, we have
%\vspace{-0.25cm}
\begin{equation}
\begin{aligned}
      	\rho(\mathbf{c})(Pj+k)=&\rho(\mathbf{a})(j)\rho(\mathbf{b})(k)+\rho(\mathbf{a})(j+1)\rho(\mathbf{b})(-P+k),
\end{aligned}
\end{equation}
where $0\le k < P$ and  $-N<  j< N$.
\end{IEEEproof}
Every row of the Hadamard matrix is mutually orthogonal to each other such that each element is a uni-modular. Since the Hadamard matrix is a square matrix, the number of mutual orthogonal sequences is equal to the length of the sequence. Further method to construct such mutually orthogonal sequences is also suggested in \cite{kumar2023direct}.
\subsection{Peak-to-mean envelope power ratio (PMEPR)}
One can model the orthogonal frequency-division multiplexing (OFDM) signal for a complex-valued word $\mathbf{a}=(a_0,a_1,\hdots,a_{N-1})$ of length $N$ as the real part of the
\begin{equation}
	s_{\mathbf{a}}(t)=\sum_{k=0}^{N-1}\zeta^{a_k+qf_kt}.
\end{equation}
where the frequency of the $i$-th sub-carrier is $f_i=f_0+i\Delta f$ for $0\le i \le K-1$, where $f_0$ and $\Delta f$ denote the carrier frequency and the sub-carrier spacing, respectively. The sequence $\mathbf{a}=(a_1,a_2,\ldots,a_N)$ is termed as modulating code-word of the OFDM signal. The PMEPR of signal $\mathbf{a}$ is defined as
\begin{equation}
    PMEPR(\mathbf{a})= \frac{1}{N}\sup_{0\leq\tau<1}{\lvert s_{\mathbf{a}}(\tau)\rvert^2}.
\end{equation}
\begin{defn}[\cite{Davis}]
	The PMEPR of transmitted signal $s_{\mathbf{a}}(t)$ is defined as
	\begin{equation}
		\text{PMEPR}(\mathbf{a})=\frac{1}{N} \sup_{0\le t<1/\Delta f}|s_{\mathbf{a}}(t)|^2.
	\end{equation}
\end{defn}
\begin{lem}[\cite{Davis}]\label{PMEPR}
	Let $\mathbf{a}$ be a sequence of length $N$. Hence, the PMEPR of transmitted signal $s_{\mathbf{a}}(t)$ satisfies
	\begin{equation}
		\text{PMEPR}(\mathbf{a})\le \frac{1}{N}\sum_{\tau=-(N-1)}^{N-1}|\rho(\mathbf{a})(\tau)|.
	\end{equation}
\end{lem}
PMEPR of the OFDM signal is calculated by considering the row of a CS or ZCS, and PMEPR of the MC-CDMA signal is calculated by the column of a CS or ZCS. The maximum over the obtained PMEPR of the row or column sequence is known as the row or column sequence PMEPR upper bound of a code, respectively.
%\vspace{-0.5cm}
\subsection{Barker sequence}
\begin{defn}
	A binary sequence $\boldsymbol{\gamma}=(\gamma_0,\gamma_1,\ldots , \gamma_{N-1})$ of length $N$ is said to be a Barker sequence  \cite{Barker}, if the AACF value of the sequence is either $\pm1$ or $0$, i.e., $|\rho(\boldsymbol{\gamma})(\tau)|$ is either $1$ or $0$ for $\tau \ne 0$.
\end{defn}
The only lengths $N>1$, for which a binary Barker sequence, $\boldsymbol{\gamma}$ is known to exist, are $2, 3, 4, 5, 7,11$ and $13$. Under one or more of the following transformations, these Barker sequences are unique for their length
\begin{equation}\label{tran}
    \gamma_i \mapsto \{-\gamma_i,\gamma_{N-1-i},(-1)^i\gamma_i\}
\end{equation}
From \textit{Lemma} \ref{PMEPR},  the PMEPR upper bound of a binary Barker sequence of odd length $p$ is $\frac{2p-1}{p}$. From \textit{Lemma} \ref{clg} and \ref{PMEPR} the PMEPR upper bound of a sequence, obtained by Kronecker product of two binary Barker sequences of odd lengths $p_1$ and $p_2$, is $\frac{(2p_1-1)(2p_2-1)}{p_1p_2}$.
\begin{table}[!ht]
		\centering
		\tiny
			\caption{List of some available CCC}\label{Tab:CCC}
		\begin{tabular}{ |c|c|c|c| }
			\hline
			 Source & CCC & constraints& element \\
	\hline
         \cite{Rat}& $(2^{k+1},2^{k+1},2^m)$ & $0<k<m$ &$q$-ary, $\frac{q}{2}\in \mathbb{N}$
 \\
 & & and $k,m\in \mathbb{Z}$ &\\
 \hline
         \cite{Praveen}& $(2^{k+1},2^{k+1},10\cdot 2^m)$ & $0<k<m$ & $q$-ary, $\frac{q}{2}\in \mathbb{N}$\\
         &  $(2^{k+1},2^{k+1},26\cdot 2^m)$ & $k,m\in \mathbb{Z}$ & \\
         \hline

         \cite{Palash_ar}& $(K,K,p_1^{m_1}p_2^{m_2}\cdots p_k^{m_k})$ & $ p_i$ is a prime,& $\rho$-ary,\\
         & & $K=p_1p_2\cdots p_k$&  $\rho$ is LCM of all $p_i$\\
        
         \hline
         \cite{Yu} & $(4,4,N)$ & $N=3,5,6,7~\&~9$ & Binary\\ 
         \hline
			\end{tabular}
	\end{table}
\section{Proposed Construction of Type-II ZCCS}\label{sec:prop}
In this section, we present our proposed type-II ZCCS construction utilizing Kronecker products. We also explore methods to reduce a code's PMEPR.
\begin{theorem}\label{main}
	Let $\mathbf{C}=\{C_1,C_2,\ldots,C_{K}\}$ be a $(K,K,N)$-CCC, where $C_i$ is a code of order $K\times N$ and $\mathbf{b}$ be a uni-modular sequence of length $P$. Then $\mathbf{Z}=\{Z_1,Z_2,\ldots,Z_{K}\}$ is a type-II $(K,K,NP-P+1,NP)$-ZCCS, where $Z_i=C_i\otimes\mathbf{b}$. Moreover, $\rho(Z_i)(\tau_1)=\rho(C_i)(0)\rho(\mathbf{b})(\tau_1)$, for $\tau_1 \le P-1$.
\end{theorem}
\begin{IEEEproof}
Each code of $\mathbf{C}$ follows the property that $\rho(C_i)(\tau)=0$ for non-zero value of $\tau$ and $\rho(C_i,C_j)(\tau^\prime)=0$ for any value of $\tau^\prime$ for $i,j\in\{0,1,\ldots,K-1\}$. 
Let 
$$C_i=[\mathbf{c}_0^i ~ \mathbf{c}_1^i~\ldots~\mathbf{c}_{K-1}^{i}]^{T},$$
then we define 
$$Z_i=C_i\otimes \mathbf{b}=[\mathbf{c}_1^i\otimes \mathbf{b} ~ \mathbf{c}_2^i\otimes \mathbf{b}~\ldots~\mathbf{c}_{K-1}^i\otimes \mathbf{b}]^{T}.$$
Since $\sum_{k=1}^{K}\rho(\mathbf{c}_k^i)(\tau)=0$ for any non-zero value of $\tau$, for $u\ge P-1$, from \textit{Lemma} \ref{clg}, it is straight forward that $\sum_{k=1}^{K}\rho(\mathbf{c}_k^i\otimes\mathbf{b})(u)=0$, i.e., $\rho(Z_i)(u)=0$ for $\lvert u\rvert \ge P-1$.\\
Since $\sum_{k=1}^{K}\rho(\mathbf{c}_k^i,\mathbf{c}_k^j)(\tau^\prime)=0$ for any integer $\tau^\prime$, from \textit{Lemma} \ref{clg}, it is straight forward that $\sum_{k=1}^{K}\rho(\mathbf{c}_k^i\otimes\mathbf{b},\mathbf{c}_k^j\otimes\mathbf{b})(u)=0$, i.e., $\rho(Z_i,Z_j)(u)=0$ for any integral value of $u$. Therefore,
%\vspace{-0.35cm}
\begin{equation*}
    \rho(Z_i,Z_j)(u)=\left\{\begin{array}{ll}
			0, & P-1<|u|<NP, i=j, \\
			0, & |u|<NP, i \neq j.
		\end{array}\right.
\end{equation*}
\end{IEEEproof}
\begin{remark}
Let $C$ be a CS of size $M\times N$ and $\mathbf{b}$ be a sequence of length $P$. From \textit{Theorem} \ref{main}, $Z=C\otimes \mathbf{b}$ be a type-II ZCS of size $M\times NP$ with ZCZ width $NP-P+1$.
\end{remark}
\begin{remark}\label{GCP_ZCP}
    Let ($\mathbf{s},\mathbf{t}$) be a GCP of length $N$ and $\mathbf{b}$ be a sequence of length $P$. From \textit{Theorem} \ref{main}, ($\mathbf{s}\otimes\mathbf{b},\mathbf{t}\otimes\mathbf{b}$) be a type-II ZCP of length $NP$ with ZCZ width $NP-P+1$. Moreover, AACF value  outside ZCZ width $\tau<P$ is $2N\rho (\mathbf{b})(\tau)$. From \eqref{PMEPR}, PMEPR upper bound is dependent on AACF values. Let PMERP of $\mathbf{b}=\phi$, then PMEPR of type-II ZCP ($\mathbf{s}\otimes\mathbf{b},\mathbf{t}\otimes\mathbf{b}$) be $2\phi$.
\end{remark}
\begin{remark}\label{Bar}
	If $\mathbf{b}$ is a Barker sequence then AACF value for any code generated by \textit{Theorem} \ref{main}, is either $\pm KN$ or $0$, outside the ZCZ width.
\end{remark}
Now, if we replace $\mathbf{b}$ by Barker sequence in \textit{Theorem} \ref{main}, then the AACF value of $Z_i$ can be reduced outside of the ZCZ width. However, as the Barker sequences are limited in length, the Kronecker product of Barker sequences is being used. According to \textit{Lemma} \ref{clg}, the Kronecker product of Barker sequences likewise has a low AACF value compared to any random choice of sequence.

The transformation mentioned in \textit{Lemma} \ref{pac} and \textit{Lemma} \ref{mac} keeps AACF value unchanged for a sequence. Multiplying a uni-modular constant to a row of code does not affect the AACF value. Furthermore, multiplying a uni-modular constant by a row of each code does not affect the ACCF value between codes. Therefore, we can reduce the column sequence PMEPR of a code with a suitable choice of uni-modular constant multiplication. One such case is in the \textit{Remark} \ref{polred}.
\begin{remark}\label{polred}
Consider $\mathbf{C}$ be a $p$-ary (CCC or ZCCS) code set such that each column can be determined by one of the transformations given in \textit{Lemma} \ref{pac} and \textit{Lemma} \ref{mac} and column sequence PMEPR is $M$ equivalent to the length of column sequence. Let $\mathbf{a}$ be a sequence of length $M$ with column PMEPR less than $\delta < M$. We multiply $i$th element of $\mathbf{a}$ to the $i$-th row of each code to reduce the column PMEPR of the code set by $\delta$. Such CCC and ZCCS can be obtained by \cite{Palash_ar, kumar2023direct}. We provide some uni modular sequences with lower PMEPR in Table \ref{Tab:uni_PMERP} other than binary Barker sequences, obtained by computer search.
\begin{table}[!ht]
		\centering
			\caption{$p$ary unimodular sequence with lower PMEPR}\label{Tab:uni_PMERP}
		\begin{tabular}{ |c|c|c|c| }
			\hline
        Length &  sequence  & $\zeta$ & PMEPR \\
        \hline
          $3$ & $(\zeta,\zeta^2, \zeta^2)$ & $e^{\frac{2\pi \sqrt{-1}}{3}}$ & $7/3$\\
	\hline
          $5$ & $(\zeta^2,\zeta^2, \zeta, \zeta^4, \zeta)$ & $e^{\frac{2\pi \sqrt{-1}}{5}}$ & $	\approx 2.294$\\
          \hline
          $6$ & $(\zeta^4,\zeta^4,\zeta^2,\zeta^5,\zeta,\zeta)$ & $e^{\frac{2\pi \sqrt{-1}}{6}}$ & $2$\\
          \hline
          $7$ & $(\zeta^5,\zeta^2,\zeta^4,\zeta,\zeta,\zeta^3,\zeta^3)$ & $e^{\frac{2\pi \sqrt{-1}}{7}}$ & $\approx 1.9765 $ \\
          \hline
			\end{tabular}
	\end{table}
\end{remark}
The CCC obtained from \cite{Palash_ar} and the ZCCS obtained from \cite{kumar2023direct} with code size $3,5,6$ and $7$ provide column PMEPR $3,5,6$ and $7$, respectively. And using \textit{Remark} \ref{polred} on the same CCC and ZCCS the column PMEPR upper bound decreases as PMEPR indicated in PMEPR column of Table \ref{Tab:uni_PMERP}.

Let $\mathcal{B}=\{\mathbf{b_1},\mathbf{b_2},\ldots,\mathbf{b_r}\}$ be a set of sequences such that $\rho(\mathbf{b_i},\mathbf{b_j})(0)=0$ for $i\ne j$, i.e., $\mathbf{b_i}\cdot \mathbf{b_j}^*=0$, also known as mutually orthogonal sequences.
\begin{theorem}\label{ZCCS_Large_set}
Let $\mathbf{C}=\{C_1,C_2,\ldots,C_K\}$ be a $(K,K,N)$-CCC and $\mathcal{B}=\{\mathbf{b_1},\mathbf{b_2},\ldots,\mathbf{b_r}\}$ be a set of mutually orthogonal sequences of length $P$, i.e., $\mathbf{b_p}\cdot \mathbf{b_{p'}^*}=0$, for $\mathbf{p}\ne \mathbf{p'}$. Then $\mathbf{Z}=\{C_1\otimes\mathbf{b_1},C_2\otimes\mathbf{b_1},\ldots,C_K\otimes\mathbf{b_1},C_1\otimes\mathbf{b_2},C_2\otimes\mathbf{b_2},\ldots,C_K\otimes\mathbf{b_2},\ldots,C_1\otimes\mathbf{b_r},C_2\otimes\mathbf{b_r},\ldots,C_K\otimes\mathbf{b_r}\}$ is a type-II $(rK,K,NP-P+1,NP)$-ZCCS.
\end{theorem}
\begin{IEEEproof}From \textit{Theorem} \ref{main}, $\rho(Z_i)(\tau)=0$, for $\lvert \tau \rvert \ge P$. Now, we partition the set $\mathbf{Z}=\{\mathbf{Z^1},\mathbf{Z^2},\ldots,\mathbf{Z^r}\}$, where
$\mathbf{Z}^p=\{C_1\otimes\mathbf{b_p},C_2\otimes\mathbf{b_p},\ldots,C_K\otimes\mathbf{b_p}\}$, for $p=1,2,\ldots,r$. From \textit{Theorem} \ref{main}, the ACCF value between any two codes from $\mathbf{Z^p}$ is zero for all time shifts. Let $Z_{\alpha}=C_{k_1}\otimes \mathbf{b_p}$ and $Z_{\beta}=C_{k_2} \otimes \mathbf{b_{p'}}$ be any two codes from $\mathbf{Z}$. From \textit{Lemma} \ref{clg},
 %\vspace{-0.25cm}
\begin{equation}\label{ex_cor_rel}
    \begin{aligned}
              \rho(Z_{\alpha},Z_{\beta})(Pj+k)&=\rho({C_{k_1},C_{k_2}})(j)\rho(\mathbf{b_p},\mathbf{b_p'})(k)\\
              &+\rho(C_{k_1},C_{k_2})(j+1)\rho(\mathbf{b_p},\mathbf{b_p'})(-P+k),
    \end{aligned}
\end{equation}
$\text{for } 0\le k < P \text{ and } -N<  j< N$.

	For any $k_1,k_2$ and $u\ne 0$, $\rho({C_{k_1},C_{k_2}})(u)=0$ which implies that  $\rho(Z_{\alpha},Z_{\beta})(\tau)=0$, for $\lvert \tau \rvert \ge P$. For $\tau=0$, $k=j=0$ and we also have $\rho(\mathbf{b_p},\mathbf{b_p'})(0)=0$. Therefore, from \eqref{ex_cor_rel}, $\rho(\mathbf{b_p},\mathbf{b_p'})(0)=0$, implying, $\rho(Z_{\alpha},Z_{\beta})(0)=0$. This completes the proof.
\end{IEEEproof}
\begin{remark}
   When $\mathcal{B}$, in \textit{Theorem} \ref{ZCCS_Large_set}, is a set of rows of a Hadamard matrix $r=P$, i.e., in this scenario it follows the bound given in \textit{Conjecture} \ref{Conj:code}.
\end{remark}
An example of the proposed type-II ZCCS is presented below.
\begin{example}\label{CCCMZ}
Let $\mathbf{C}=\{C_1,C_2,C_3,C_4\}$ is a binary $(4,4,8)$-CCC given in \textit{Table} \ref{C}, where $+$ and $-$ represent $1$ and $-1$, respectively.\\
\begin{table}[!ht]
		\centering
		%\tiny
	\begin{tabular}{|c|c|c|c|}
			\hline
		%	$C_1$ & $C_2$ & $C_3$ & $C_4$\\
		%	\hline
			$\begin{smallmatrix}
    -     +     +     +     +     +     +    -\\
    -    -     +    -     +    -     +     +\\
    -     +    -    -     +     +    -     +\\
     +     +     +    -    -     +     +     +

			\end{smallmatrix}$
			&
		$\begin{smallmatrix}
    -    -     +    -     +    -     +     +\\
    -     +     +     +     +     +     +    -\\
    -    -    -     +     +    -    -    -\\
     +    -     +     +    -    -     +    -
		\end{smallmatrix}$
        &
$\begin{smallmatrix}
    -     +    -    -     +     +    -     +\\
     +     +     +    -    -     +     +     +\\
    -     +     +     +     +     +     +    -\\
    -    -     +    -     +    -     +     +

\end{smallmatrix}$
			&
$\begin{smallmatrix}
     +     +     +    -    -     +     +     +\\
    -     +    -    -     +     +    -     +\\
     +     +    -     +    -     +    -    -\\
     +    -    -    -    -    -    -     + 
\end{smallmatrix}$ \\
			\hline
			\end{tabular} 
			\caption{Binary CCC}\label{C}
	\end{table}
Let $\mathbf{b}^1=(+,+)$ and $\mathbf{b}^1=(+,-)$ be mutually orthogonal sequence of length $2$. Using \textit{Theorem} \ref{ZCCS_Large_set}, binary type-II $(8,4,15,16)$-ZCCS can be obtained as given in the \textit{Table} \ref{ZCCS}. AACF value of each code from $\mathbf{Z}$ is $(0_{14},32,64,32,0_{14})$ and AACF value of any two code from $\mathbf{Z}$ is either $(0_{14},32,0,32,0_{14})$ or $(0_{31})$. AACF value value of each column sequence is either $(1,0,-1,0,4,0,-1,0,1)$ or $(-1,0,1,0,4,0,1,0,-1)$. Therefore, the column sequence PMEPR is upper bounded by $2$ for this given example.
%\vspace{-0.25cm}
\begin{table}[!ht]
		\centering
		%\tiny 
		\begin{tabular}{|c|c|}
			\hline
			$Z_1$ & $Z_2$ \\
			\hline
			$\begin{smallmatrix}
     -    -     +     +     +     +     +     +     +     +     +     +     +     +    -    -\\
    -    -    -    -     +     +    -    -     +     +    -    -     +     +     +     +\\
    -    -     +     +    -    -    -    -     +     +     +     +    -    -     +     +\\
     +     +     +     +     +     +    -    -    -    -     +     +     +     +     +     +
\end{smallmatrix}$ &
			%	 \\
		%	\hline
			$\begin{smallmatrix}
       -     +     +    -     +    -     +    -     +    -     +    -     +    -    -     +\\
    -     +    -     +     +    -    -     +     +    -    -     +     +    -     +    -\\
    -     +     +    -    -     +    -     +     +    -     +    -    -     +     +    -\\
     +    -     +    -     +    -    -     +    -     +     +    -     +    -     +    -
\end{smallmatrix}$\\
\hline
$Z_3$ &$Z_4$ \\
			\hline
			$\begin{smallmatrix}
     -    -    -    -     +     +    -    -     +     +    -    -     +     +     +     +\\
    -    -     +     +     +     +     +     +     +     +     +     +     +     +    -    -\\
    -    -    -    -    -    -     +     +     +     +    -    -    -    -    -    -\\
     +     +    -    -     +     +     +     +    -    -    -    -     +     +    -    -

\end{smallmatrix}$ &
%  \\
%			\hline
			$\begin{smallmatrix}
     -     +    -     +     +    -    -     +     +    -    -     +     +    -     +    -\\
    -     +     +    -     +    -     +    -     +    -     +    -     +    -    -     +\\
    -     +    -     +    -     +     +    -     +    -    -     +    -     +    -     +\\
     +    -    -     +     +    -     +    -    -     +    -     +     +    -    -     +
\end{smallmatrix}$\\
\hline
$Z_5$ &$Z_6$ \\
			\hline
			$\begin{smallmatrix}
    -    -     +     +    -    -    -    -     +     +     +     +    -    -     +     +\\
     +     +     +     +     +     +    -    -    -    -     +     +     +     +     +     +\\
    -    -     +     +     +     +     +     +     +     +     +     +     +     +    -    -\\
    -    -    -    -     +     +    -    -     +     +    -    -     +     +     +     +
\end{smallmatrix}$ &
%  \\
%			\hline
			$\begin{smallmatrix}
    -     +     +    -    -     +    -     +     +    -     +    -    -     +     +    -\\
     +    -     +    -     +    -    -     +    -     +     +    -     +    -     +    -\\
    -     +     +    -     +    -     +    -     +    -     +    -     +    -    -     +\\
    -     +    -     +     +    -    -     +     +    -    -     +     +    -     +    -
\end{smallmatrix}$\\
\hline
$Z_7$ &$Z_8$ \\
			\hline
			$\begin{smallmatrix}
     +     +     +     +     +     +    -    -    -    -     +     +     +     +     +     +\\
    -    -     +     +    -    -    -    -     +     +     +     +    -    -     +     +\\
     +     +     +     +    -    -     +     +    -    -     +     +    -    -    -    -\\
     +     +    -    -    -    -    -    -    -    -    -    -    -    -     +     +
\end{smallmatrix}$ &
			$\begin{smallmatrix}
     +    -     +    -     +    -    -     +    -     +     +    -     +    -     +    -\\
    -     +     +    -    -     +    -     +     +    -     +    -    -     +     +    -\\
     +    -     +    -    -     +     +    -    -     +     +    -    -     +    -     +\\
     +    -    -     +    -     +    -     +    -     +    -     +    -     +     +    - 
\end{smallmatrix}$\\
\hline
\end{tabular}
\caption{Binary type-II ZCCS.}\label{ZCCS}
	\end{table}
\end{example}
\begin{theorem}\label{Th:itr_ZCCS}
Let $\mathbf{B}=\{B_1,B_2,\ldots,B_{K_1}\}$ be a type-II $(K_1,M_1,Z,N_1)$-ZCCS and $\mathbf{C}=\{C_1,C_2,$ $\ldots,C_{K_2}\}$ be a $(K_2,K_2,N_2)$-CCC. Then $\mathbf{Z}=\{C_1\otimes B_1,C_1\otimes B_2,\ldots, C_1\otimes B_{K_1},C_2\otimes B_1,C_2\otimes B_2,\ldots, C_2\otimes B_{K_1},\ldots, C_{K_2}\otimes B_1,C_{K_2}\otimes B_2,\ldots, C_{K_2}\otimes B_{K_1}\}$, is a type-II $(K_1K_2,M_1K_2,N_1N_2-N_1+Z,N_1N_2)$ ZCCS.
\end{theorem}
\begin{IEEEproof}
Let $Z_{(u-1)K2+v}=C_u\otimes B_v$, for $1\le u \le K_1$ and $1\le v\le K_2$. Now we choose $1\le k_1,k_2\le K_1K_2$, such that $k_1=(u_1-1)K_2+v_1$ and $k_2=(u_2-1)K_2+v_2$. From \textit{Lemma} \ref{clg},
    \begin{equation}
    \begin{aligned}
        \rho(Z_{k_1},Z_{k_2})(N_1j+k)=&\rho(C_{u_1},C_{u_2})(j) \rho(B_{v_1},B_{v_2})(k)\\
        &+\rho(C_{u_1},C_{u_2})(j+1) \rho(B_{v_1},B_{v_2})(-N_1+k).
        \end{aligned}
    \end{equation}
    Since $\mathbf{B}$ is a  type-II $(K_1,M_1,Z,N_1)$-ZCCS, $\rho(B_{v_1},B_{v_2})(k)=0$, whenever $k\ge N_1-Z.$ Therefore, it can be easily seen that $\rho(Z_{k_1},Z_{k_2})(\tau)=0$, whenever $\tau\ge N_1-Z $. Let $k_1\ne k_2$, in this scenario we have $u_1\ne u_2$ or $v_1\ne v_2$.
    \begin{equation}\label{Eq:corr_zero}
        \rho(Z_{k_1},Z_{k_2})(0)=\rho(C_{u_1},C_{u_2})(0) \rho(B_{v_1},B_{v_2})(0).
    \end{equation}
    Since, at least one of $u_1,u_2$ or $v_1,v_2$ not same, which makes right hand side of \eqref{Eq:corr_zero} is zero, i.e., $\rho(Z_{k_1},Z_{k_2})(0)=0$, whenever $k_1\ne k_2$. This completes the proof.
\end{IEEEproof}
\begin{remark}
Let $\mathbf{B}$ in \textit{Theorem} \ref{Th:itr_ZCCS} satisfy the number of codes relation given in \textit{Conjecture} \ref{Conj:code}, then the obtained type-II ZCCS also follows the number of codes relation given in \textit{Conjecture} \ref{Conj:code}.
\end{remark}

 The number of codes for a type-I $(K,M,Z,N)$-ZCCS is bounded by $K \leq M \left\lfloor \frac{N}{Z}\right\rfloor$ \cite{Fan}. The number of codes for the proposed type-II $(K, M, Z, N)$-ZCCS is bounded by $K \le M (N - Z + 1)$, which is larger than that of type-I ZCCS.

Now we are providing a type-I ZCCS with $8$ codes having $4$ constituent sequence of length $16$. Due to $K \leq M \left\lfloor \frac{N}{Z}\right\rfloor$, $Z\le 8$. Type-I ZCCS given in TABLE \ref{T1ZCCS} have AACF value of each code is $(0_7,32,0_7,64,0_7,32,0_7)$ and ACCF value between two codes are either $(0_7,32,0_7,64,0_7,32,0_7)$ or $(0_{31})$.
\begin{table}[!ht]
		\centering
		%\tiny 
		\begin{tabular}{|c|c|}
			\hline
			$A_1$ & $A_2$ \\
			\hline
			$\begin{smallmatrix}
    -     +     +     +     +     +     +    -    -     +     +     +     +     +     +    -\\
    -    -     +    -     +    -     +     +    -    -     +    -     +    -     +     +\\
    -     +    -    -     +     +    -     +    -     +    -    -     +     +    -     +\\
     +     +     +    -    -     +     +     +     +     +     +    -    -     +     +     +
\end{smallmatrix}$ &
			%	 \\
		%	\hline
			$\begin{smallmatrix}
    -     +     +     +     +     +     +    -     +    -    -    -    -    -    -     +\\
    -    -     +    -     +    -     +     +     +     +    -     +    -     +    -    -\\
    -     +    -    -     +     +    -     +     +    -     +     +    -    -     +    -\\
     +     +     +    -    -     +     +     +    -    -    -     +     +    -    -    -
\end{smallmatrix}$\\
\hline
$A_3$ &$A_4$ \\
			\hline
			$\begin{smallmatrix}
     -    -     +    -     +    -     +     +    -    -     +    -     +    -     +     +\\
    -     +     +     +     +     +     +    -    -     +     +     +     +     +     +    -\\
    -    -    -     +     +    -    -    -    -    -    -     +     +    -    -    -\\
     +    -     +     +    -    -     +    -     +    -     +     +    -    -     +    -
\end{smallmatrix}$ &
%  \\
%			\hline
			$\begin{smallmatrix}
    -    -     +    -     +    -     +     +     +     +    -     +    -     +    -    -\\
    -     +     +     +     +     +     +    -     +    -    -    -    -    -    -     +\\
    -    -    -     +     +    -    -    -     +     +     +    -    -     +     +     +\\
     +    -     +     +    -    -     +    -    -     +    -    -     +     +    -     +
\end{smallmatrix}$\\
\hline
$A_5$ &$A_6$ \\
			\hline
			$\begin{smallmatrix}
    -     +    -    -     +     +    -     +    -     +    -    -     +     +    -     +\\
     +     +     +    -    -     +     +     +     +     +     +    -    -     +     +     +\\
    -     +     +     +     +     +     +    -    -     +     +     +     +     +     +    -\\
    -    -     +    -     +    -     +     +    -    -     +    -     +    -     +     +
\end{smallmatrix}$ &
%  \\
%			\hline
			$\begin{smallmatrix}
     -     +    -    -     +     +    -     +     +    -     +     +    -    -     +    -\\
     +     +     +    -    -     +     +     +    -    -    -     +     +    -    -    -\\
    -     +     +     +     +     +     +    -     +    -    -    -    -    -    -     +\\
    -    -     +    -     +    -     +     +     +     +    -     +    -     +    -    -
\end{smallmatrix}$\\
\hline
$A_7$ &$A_8$ \\
			\hline
			$\begin{smallmatrix}
     +     +     +    -    -     +     +     +     +     +     +    -    -     +     +     +\\
    -     +    -    -     +     +    -     +    -     +    -    -     +     +    -     +\\
     +     +    -     +    -     +    -    -     +     +    -     +    -     +    -    -\\
     +    -    -    -    -    -    -     +     +    -    -    -    -    -    -     +
\end{smallmatrix}$ &
			$\begin{smallmatrix}
     +     +     +    -    -     +     +     +    -    -    -     +     +    -    -    -\\
    -     +    -    -     +     +    -     +     +    -     +     +    -    -     +    -\\
     +     +    -     +    -     +    -    -    -    -     +    -     +    -     +     +\\
     +    -    -    -    -    -    -     +    -     +     +     +     +     +     +    -
\end{smallmatrix}$\\
\hline
\end{tabular}
\caption{Binary type-I ZCCS.}\label{T1ZCCS}
	\end{table}
\section{System Model and Performance Analysis}\label{sec:system}
\begin{figure}
				\centerline {
				\includegraphics[width=8cm,height=7cm]{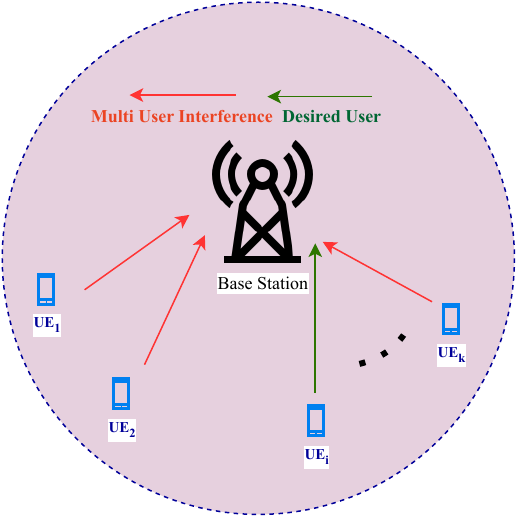}
					}
				\caption{QS-Uplink Scenario}\label{fig:uplink}	
			\end{figure}

In this section, we provide a ZCCS-based QS MC-CDMA system for QS uplink communication. Consider a single-cell SISO uplink scenario as shown in Fig.\ref{fig:uplink}, where each user equipment (UE) communicating with the base station (BS) is at a different distance from the BS, which means the signals from the different users arriving at the BS are not synchronized in time. In this ZCCS-based QS MC-CDMA system, each user is assigned a unique code matrix of dimension $M\times N$ from the proposed ZCCS set.\par
The binary phase shift keying (BPSK) modulated data bit $b^k(n)$ of each user $k$ is spread by the M element codes of their allocated code matrix $\mathbf{Z}^k(n)$ to be broadcast on M separate subcarriers. The transmitted signal intended for $m$th subcarrier of $K$th user can be given as
\begin{equation}\label{eq:skmn}
	\mathbf{s}^{k}_{m}(n)= b^{k}(n){\mathbf{z}^{k}_{m}}(n-t_k),~~m= 1,2,...,M  ,
	\end{equation}
where $t_k$ is the relative delay associated with the $k$th user such that $N-Z<|t_k|<N$ and $(t_k=0)$.  Stacking data intended for all $M$ subcarriers, we get a matrix to spread data of $k$th user with size $M\times N$, i.e.,
\begin{equation}\label{eq:skn}
	\mathbf{S}^{k}(n)=\begin{bmatrix}
(\mathbf{s}^{k}_1(n))^T, (\mathbf{s}^{k}_2(n))^T, \hdots, (\mathbf{s}^{k}_M(n))^T
\end{bmatrix}^T.
	\end{equation}
 $M$ point inverse fast Fourier transform (IFFT) of the matrix in equation \eqref{eq:skn} is done for sending the different flocks of element codes on different subcarriers. After IFFT, a cyclic prefix (CP) of length one-fourth of the number of subcarriers is added to avoid inter-symbol interference. similar operations are performed on all the $K$ users' data and relative delay between these users has been introduced to make the system quasi-synchronous.

Now, for an uplink scenario, all the signals for these $K$ users are passed through $K$ separate L-path Rayleigh fading channels which can be modelled as
\begin{equation}\label{eq:hk}
 h^k(\tau,n) = \sum_{l=0}^{L-1}\alpha_{l}^{k}(n)\delta(\tau-\tau_{l}^{k}), ~~ l = 0,1,...,L-1 , k = 1,2,...,K ,
\end{equation}
where $\alpha_{l}^{k}(.)$ is the gain of the $l$th path of $k$th user and $\tau_{l}^{k}$ is the delay associated with that path. The received signal at the receiver can be given as 
\begin{equation}\label{eq:y}
\mathbf{Y} = \sum_{k=1}^{K}\mathbf{H}^k\mathbf{S}^{k}(n)+\mathbf{N}_0,
\end{equation}
where $\mathbf{H}^k$ is the circulant channel matrix formed by $k$th user's channel impulse response and has dimension $ M\times M$ and $\mathbf{N}_0$ is the additive white Gaussian noise (AWGN) matrix. Fast Fourier transform is performed on the received matrix $\mathbf{Y}$ in \eqref{eq:y} to convert the time domain signal back to the frequency domain and CP is removed to get $\tilde{\mathbf{Y}}$. After the removal of CP channel equalization is performed by dividing the resulting matrix by the complex conjugate of the channel matrix to equalize the effect of the channel 
\begin{equation}\label{eq:r}
     \mathbf{R}= \frac{\tilde{\mathbf{Y}}}{\mathbf{H^*}}, 
\end{equation}

The $M$ rows of the resulting matrix $\mathbf{R}$ are fed to the bank of $M$ correlators tuned to the $M$ rows (each for one subcarrier frequency) of the code matrix of the desired user. The output of the $m$th correlator can be given as 
\begin{equation}\label{eq:o}
	\mathbf{o}_{m}=\sum_{k=1}^{K}\rho(\mathbf{r}_m,\mathbf{z}^k_{m})(t_k),
	\end{equation}
 where $\mathbf{r}_m$ is the $m$th row vector of the matrix $\mathbf{R}$, $\mathbf{z}^k_{m}$ is the $m$th element sequence and $t_k$ is the relative delay of the $k$th user respectively. Considering user 1 as a desired user without loss of generality $t_1=0$ and performing summation over subcarriers the decision variable can be given as 
\begin{equation}\label{eq:d}
\begin{aligned}
\mathbf{d}^1 = & \underbrace{ b^1(n)\sum_{m=1}^{M}\rho(\mathbf{z}^1_m(n),\mathbf{z}^1_m(n))(0)}_{\text{Desired data}}+\underbrace{b^1(n)\sum_{m=1}^{M}\sum_{l=1}^{L-1}\rho(\mathbf{z}^1_m(n),\mathbf{z}^1_m(n))(\tau^1_l)}_{\text{Multi path interference}} \\
&+\underbrace{b^{k(0)}(n)\sum_{k=2}^{K}\sum_{m=1}^{M}\rho(\mathbf{z}^k_m(n),\mathbf{z}^1_m(n))(t_k)+b^{k(-1)}(n)\sum_{k=2}^{K}\sum_{m=1}^{M}\rho(\mathbf{z}^k_m(n),\mathbf{z}^1_m(n))(N-t_k)}_{\text{Multi access interference}}\\
&+\underbrace{\sum_{k=1}^{K}\sum_{m=1}^{M}\mathbf{n}_{m}\rho(\mathbf{z}^k_m(n),\mathbf{z}^k_{m}(n)(t_k)}_{\text{Noise}},    
\end{aligned}
\end{equation}
where $b^{k(0)}, b^{k(-1)}$ is the current and previous bit of the $k$th user respectively and $N$ is the length of an element code. From the definition of type-II ZCCS in \eqref{eq:def_ZCCS} the MPI and MAI in the \eqref{eq:d} becomes completely zero if $N-Z<|\tau^1_l,t_k|<N$ and $t_k=0$.

    \begin{figure}
				\centerline {
				\includegraphics[width=16cm,height=7cm]{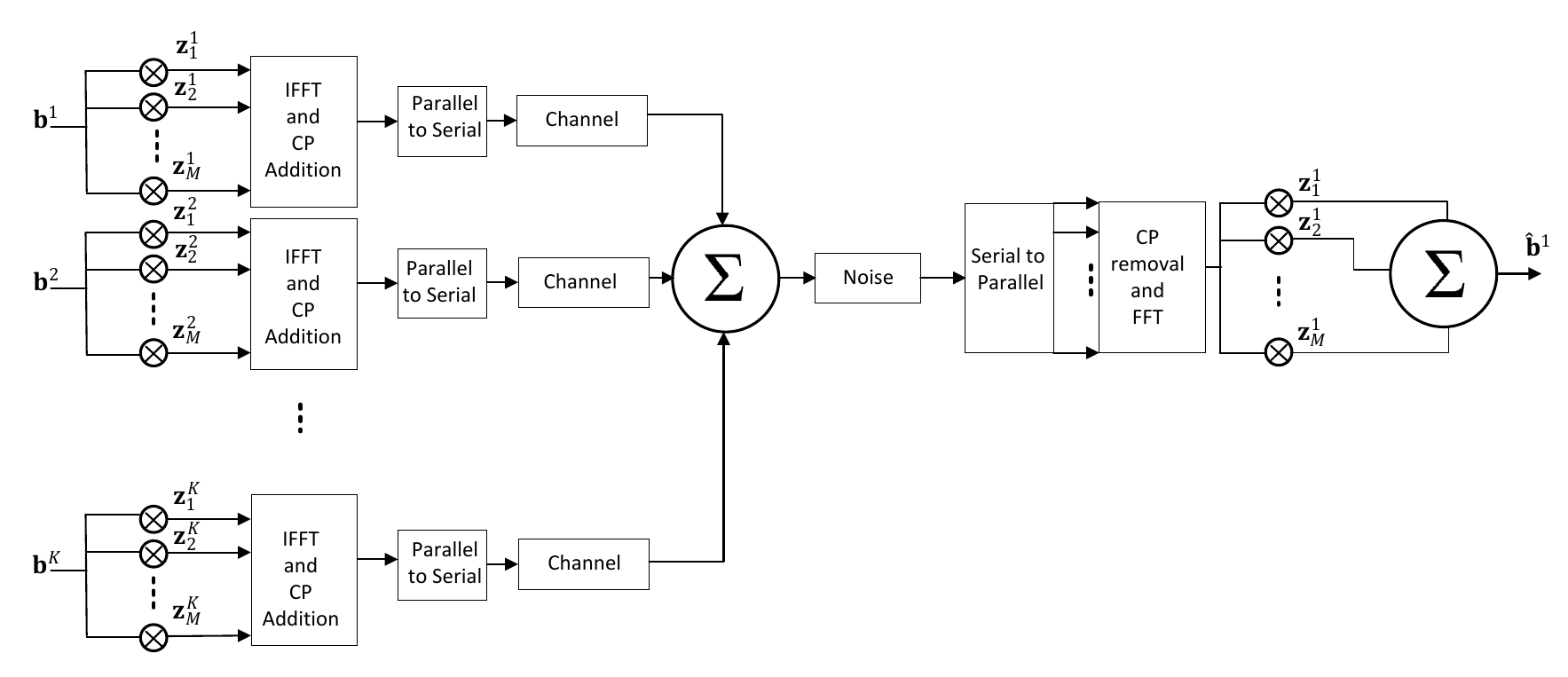}
					}
				\caption{Uplink QS MC-CDMA system model }\label{fig:system_model}	
			\end{figure}
\begin{figure}[!ht]
				\centerline {
				\includegraphics[width=8cm,height=7cm]{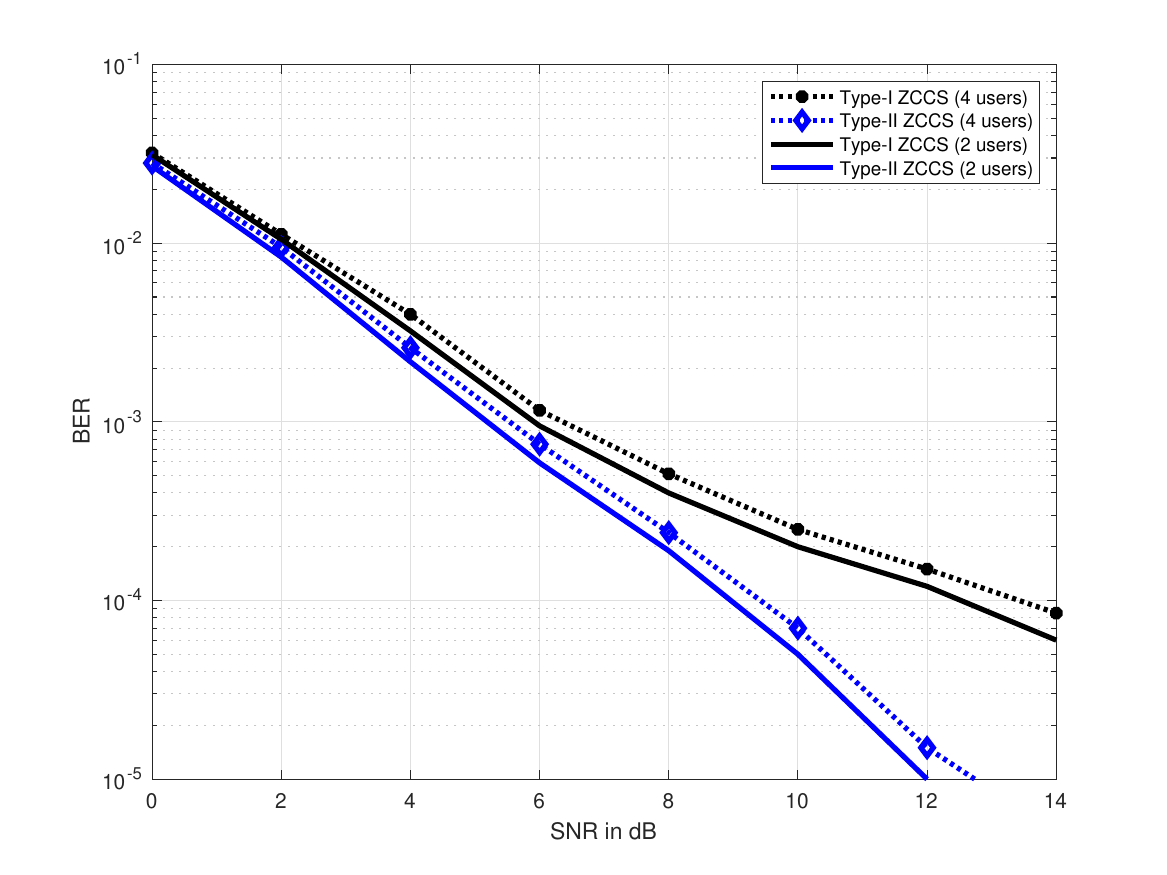}
					}
				\caption{BER performance comparison of uplink QS MC-CDMA in Rayleigh fading environment. }\label{fig:comp_BER}	
			\end{figure}

A MATLAB simulation was performed to evaluate and compare the performance of the proposed code with an existing type I ZCCS. In this study, we have examined a Single Input Single Output (SISO) uplink Multi-Carrier Code Division Multiple Access (MC-CDMA) scenario inside a multipath Rayleigh fading environment. The simulation was conducted with respect to a system including two users and another system comprising four users. The Rayleigh channels under consideration possess four taps. For spreding type-II $(8,4,15,16)$-ZCCS and type-I $(8,4,7,16)$-ZCCS has been considered. The simulation has been performed on $10^5$ symbols for $500$ Monte-Carlo iterations.The BER performance of the multiuser QS MC-CDMA system is demonstrated across various signal-to-noise ratio (SNR) values in Fig.\ref{fig:comp_BER}.
   
The performance of the system with two users is shown by solid lines, whereas the performance of the same system with four users is represented by dotted lines with markers. The figure provides strong evidence that the proposed code set exhibits superior performance compared to the existing type-I ZCCS in both cases. This may be attributed to the bigger zero cross-correlation zone of the proposed code. It is noteworthy that the disparity in performance between systems accommodating two users and systems accommodating four users in a QS uplink scenario is minimal. Another benefit of the proposed code set in comparison to the existing code set is its far higher delay tolerance, which is nearly twice as much. Additionally, More users can be accommodated with the same flock size and ZCZ width as compared to type-I ZCCS.  

\section{Comparison with Existing work}\label{sec:comp}
\subsection{Comparison with Existing Type-II ZCP}
The proposed construction also features type-II ZCP of length in the form of $NP$, where $P$ is any natural number and  $N$ is the length of available GCPs. Thus type-II ZCP can be constructed of lengths in the form of $N,2N,3N,$ and so on. However, \cite{Zeng_ZCP_WCL_2022} constructed quadriphase type-II ZCP with the length in the forms of $3N$, $7N$, $9N$, $14N$ and $15N$ only.
\begin{table}[!ht]
\centering
\tiny
\begin{tabular}{ |c|c|c|c| }
 \hline
 Method & type-II ZCP & AACF value & PMEPR\\
        &     & $\tau=(0,1,\ldots,23)$& \\
  \hline
    \cite{Rajen} & $\begin{matrix}
          + - + + - + + - + - + - 
          + - + + - + - + - + - +\\
          - + - + - + - + - - + -
           - + - + -  + - + + - +
          \end{matrix}$ & $(48,32,16,0_{21})$ &$8$\\
          \hline
          [Proposed]  & $\begin{matrix}
          - - + + + - + + - - - +
           + + - + + - + + - + + -\\
          - - + - - + + + - + + -
          + + - - - + + + - - - +
          \end{matrix}$&  $(48,0,16,0_{21})$ & $10/3$\\
          \hline
   \end{tabular}
\caption{Binary Type-II ZCPs of length $24$.}\label{T3}
\end{table}
\par In the proposed construction, if we consider GCPs of length in the form of power-of-two, then every length of type-II ZCP mentioned in \cite{Rajen} can obtain with the same ZCZ width. As we use GCPs of every possible length, our proposed construction provides more lengths and large ZCZ width with a low AACF value outside the ZCZ width. Therefore, the PMEPR of the proposed ZCP is lower than the PMEPR of ZCPs constructed in \cite{Rajen}.
\subsection{Comparison between type-I ZCCS and type-II ZCCS}

 The proposed code has good properties with larger ZCZ compared to type-I ZCCS. For sequence lengths of $N$ and ZCZ widths of $Z$, the maximum set size to flock size ratio for CCC is $1$,  for type-I ZCCS it is $\left\lfloor \frac{N}{Z}\right\rfloor$, and for type-II, it is $N-Z+1$. 
As a result, with the same flock size, the proposed set offers
a substantially larger number of codes compared to type-I ZCCS. Also from section \ref{sec:system} it is evident that the proposed ZCCS outperforms the existing ZCCS.
\section{Conclusion}\label{sec:conc}
This paper proposes a new code set, type-II ZCCS, and its construction for arbitrary sequence length with a larger set size and ZCZ width. The proposed construction also generalizes some of the existing type-II ZCP and produces type-II ZCS. The proposed type-II ZCCS provides more codes, better BER, improved PMEPR, and larger ZCZ width than type-I ZCCS.
 \bibliographystyle{IEEEtran}
 \bibliography{Rajen}
 \end{document}